\documentclass[iop]{emulateapj}
\usepackage{graphicx}
\usepackage{amsmath}
\usepackage{atbegshi}

\begin{document}

\newcommand{\cubi}{$C_{U,B,I}$ }
\newcommand{\chst}{$C_{F275W,F336W,F435W}$ }

\title{A PHOTOMETRIC STUDY OF GLOBULAR CLUSTERS OBSERVED BY THE APOGEE SURVEY}

\author{
Szabolcs~M{\'e}sz{\'a}ros\altaffilmark{1,2},
D.~A.~Garc\'{\i}a-Hern{\'a}ndez\altaffilmark{3,4}, 
Santi~Cassisi\altaffilmark{5}, 
Matteo~Monelli\altaffilmark{2,3},
L{\'a}szl{\'o}~Szigeti\altaffilmark{1}, 
Flavia~Dell'Agli\altaffilmark{3,4},
Al{\'i}z~Derekas\altaffilmark{1,7}, 
Thomas~Masseron\altaffilmark{3,4},
Matthew~Shetrone\altaffilmark{6},
Peter~Stetson\altaffilmark{8},
Olga~Zamora\altaffilmark{3,4}
}

\altaffiltext{1}{ELTE E\"otv\"os Lor\'and University, Gothard Astrophysical Observatory, Szombathely, Hungary}
\altaffiltext{2}{Premium Postdoctoral Fellow of the Hungarian Academy of Sciences}
\altaffiltext{3}{Instituto de Astrof{\'{\i}}sica de Canarias (IAC), E-38205 La Laguna, Tenerife, Spain}
\altaffiltext{4}{Universidad de La Laguna, Departamento de Astrof\'{\i}sica, 38206 La Laguna, Tenerife, Spain}
\altaffiltext{5}{INAF-Osservatorio Astronomico di Teramo, via M. Maggini, I-64100 Teramo, Italy}
\altaffiltext{6}{University of Texas at Austin, McDonald Observatory, Fort Davis, TX 79734, USA}
\altaffiltext{7}{Konkoly Observatory, MTA CSFK, Konkoly Thege Mikl\'os \'ut 15-17, H-1121 Budapest, Hungary}
\altaffiltext{8}{Herzberg Astronomy and Astrophysics, National Research Council Canada, 5071 West Saanich Road, Victoria, BC V9E 2E7, Canada}

\begin{abstract}

In this paper we describe the photometric and spectroscopic properties of multiple populations in 
seven northern globular clusters. In this study we employ precise ground based photometry from 
the private collection of Stetson, space photometry from the Hubble Space Telescope,
literature abundances of Na and O, and APOGEE survey abundances for Mg, Al, C, and N.
Multiple populations are identified by their position in the \cubi-V 
pseudo-CMD and confirmed with their chemical composition determined using abundances. We confirm the expectation
from previous studies that the RGB in all seven clusters are split and the different branches have 
different chemical compositions.  The Mg-Al anti-correlations were well explored 
by the APOGEE and Gaia-ESO surveys for most globular clusters, some clusters showing bimodal distributions, 
while others continuous distributions. Even though the structure (i.e., bimodal vs. continuous) of Mg-Al can greatly 
vary, the Al-rich and Al-poor populations do not seem to have very different photometric properties, agreeing with 
theoretical calculations. There is no one-to-one correspondence between the Mg-Al anticorrelation 
shape (bimodal vs. continuous) and the structure of the RGB seen in the HST pseudo-CMDs, with the HST
photometric information usually implying more complex formation/evolution
histories than the spectroscopic ones. We report on finding two second generation HB stars in M5, and five second 
generation AGB stars in M92, which is the most metal-poor cluster to date in which second generation AGB stars 
have been observed.

\end{abstract}

\section{Introduction}

Multiple populations in globular clusters (GCs) are well known today. They are extensively studied in 
the literature using both photometric and spectroscopic data. To date almost all GCs were found to have
multiple main sequences and/or subgiant and/or giant branches, \citep[e.g.,][]{piotto01, milone01, piotto02}, 
which are accompanied by variations in the content of He and light elements, and a small age difference between 
the distinct sub-stellar populations \citep{dantona02, cassisi01}, except in $\omega$~Cen \citep{marino06}. 
For most clusters the C+N+O content in globular clusters is fairly constant to within 0.3~dex 
\citep{ivans02, carretta04, smith03}, and only a handful of clusters have been found where this is not the 
case, like N1851 \citep{yong04}, $\omega$~Cen \citep{marino06}, M22 \citep{alves01}, and M2 \citep{lardo01, lardo03}.
The formation and evolution of GCs turned out to be a more complex problem than previously thought 
\citep{kraft01, gra01}, and no individual model is capable
of fully explaining the evolution of these objects \citep[see e.g.,][for a review]{renzini01}.

Multiple photometric surveys have had the goal to characterize GCs to understand their formation and evolution. 
The Hubble Space Telescope (HST) Treasury Project \citep{soto01, piotto02} provides the largest and most precise homogeneous photometric data set of 
photometry in five filters for 47 GCs.
The SUMO project \citep{monelli01} is a ground based, homogeneous, photometric study of 
multiple stellar populations in the largest sample of GCs. 

The different populations in each cluster have different chemical compositions, that mostly 
manifest in light element variations (O, Na, C, N, Mg and Al) along the red giant branch. These 
elements are known to (anti)-correlate with each other, and are the result of high-temperature H-burning. 
The most well studied anti-correlations are Na-O and Al-Mg. The most extensive study of the southern 
GCs were carried out by \citet{carretta02, carretta03, carretta01}, who focused on the Na-O and 
Al-Mg anti-correlations. They showed that while these anti-correlations are unique to GCs, the structure or shape 
of the anti-correlation patterns and spread of 
abundances depends on the total mass and metallicity of the clusters. 

For the northern clusters, the largest homogeneous study 
was presented by \citet{meszaros01} based on data from the Apache Point Observatory Galactic Evolution Experiment
\citep[APOGEE;][]{majewski01} part of the 3rd Sloan Digital Sky Survey \citep[SDSS-III;][]{eis11}. APOGEE was a high 
resolution near-infrared survey focused on the H-band \citep[15,090 to 16,990 \AA;][]{wil10}, and observed more than 
100,000 red giant stars from all components of the Milky Way. The survey lasted from 2011 to 2014, and its successor, APOGEE-2, 
will continue until 2020. \citet{meszaros01} were able to conduct a
more detailed analysis of the Mg-Al anti-correlation, 
because more stars with higher Al abundances were observed than in previous studies. 
This allowed the discovery of significantly different shapes in Mg-Al anti-correlation between clusters.
Analysis of the CO and CN lines in the H-band made it possible to measure [C/Fe] and [N/Fe] abundance ratios for most stars, 
revealing the C-N anti-correlation in the whole sample of investigated clusters; however, the measurement error of these 
abundances was relatively high.
For the southern clusters, the most recent examination of Mg and Al was carried out by \citet{pancino01}.
They used Gaia-ESO DR4 data to explore the Mg-Al anti-correlation in nine clusters and found extended anti-correlations 
in only the more metal-poor clusters.

Combining photometry with abundances is a powerful tool in understanding GC formation/evolution \citep{monelli01} 
and it was fundamental in discovering second generation asymptotic giant branch (SG-AGB) stars in GCs. The lack of SG-AGB 
stars in globular clusters puzzled astronomers in recent years. \citet{campbell01} did not found Na-rich AGB stars 
in NGC~6752, which is the main tracer element of second generation 
stars in globular clusters. The possible lack of SG-AGB stars presented a challenge for stellar 
evolution models and the formation of GCs \citep{charbonnel01, cassisi02}. \citet{cassisi02} argued the lack 
of SG-AGB stars in NGC~6752 is not consistent with star counts along the HB and AGB as well as with the canonical stellar 
models. Early evidence showing the contrary view came from photometry of NGC~2808 by \citet{milone03}, who observed
three different populations along the AGB. Later, \citet{johnson03} found Na-rich AGB stars in the metal-rich GC 
47~Tuc \citep[see also][]{lapenna02}. Finally, \citet{garcia03} definitely 
solved the apparent tension between observations and models by showing clear evidence of the presence of 
fourteen SG-AGB stars in four different  metal-poor ([Fe/H]$<-$1.0) clusters (M13, M5, M3, and M2). This discovery was based upon 
using Al as a tracer instead of Na to identify second generation stars. As found by \citet{meszaros01}, 
deriving [Al/Fe] from the atomic lines of Al in the H-band can be clearly used to separate second 
generation stars from first generation. Later SG-AGB stars were also found in M4 \citep{lardo02, marino04}, 
NGC~6752 \citep{lapenna01}, and NGC~2808 \citep{marino04}.

In this paper, we focus on combining the large abundance data set available from the literature with 
ground based and HST photometry in order to study the differences in the photometric properties of clusters 
with bimodal and continuous Mg-Al anti-correlations. In addition, we report the first
discovery of SG-AGB stars in M92; the  most metal-poor cluster in which such stars have been observed.

\section{Linking Spectroscopy with Photometry}

The ground based U, B, V, R, I photometry was taken from the private collection of Peter Stetson and 
is precise to the level of $<$ 0.002 mag in the U band and to $<$ 0.001 mag for the other bands \citep{stetson01}.

Photometric and APOGEE data are currently available for seven northern globular clusters. The advantage of 
APOGEE data is that all abundances were derived consistently, which allows us a more accurate direct 
comparison between clusters. Abundances of Fe, Mg, Al were derived using neutral atomic lines, that were 
believed to be less affected by NLTE effects than lines in the optical, because they are formed deeper in 
the atmosphere. However, it was shown recently that this is not the case, as corrections larger than 0.1~dex
may be needed for both Mg \citep{zhang01} and Al \citep{nord01}. 
The O abundances were derived from OH lines, then with the O abundances held constant the [C/Fe] ratios were 
determined from the CO lines. The final step 
in the process was to determine [N/Fe] from CN. The Na lines in the H-band are 
too weak for any measurements below [M/H] $< -$0.7~dex; thus, \citet{meszaros01} was unable to study the Na-O 
anti-correlation, even though O measurements were available. 

In order to extend our study to the Na-O anti-correlation, we collected abundances of both elements
from the literature (see Table 1 for a full list of references). 
The sample was limited to studies which sampled significant part of the RGB with 
many stars observed, so only two clusters are discussed in detail, M13 and M5. \citet{carretta03}  
obtained the largest set of Na and O abundances for the most clusters to date, and discussed M5 in detail. 
The spectra were acquired with FLAMES/GIRAFFE mounted on the VLT UT2, and they used the forbidden O 
lines at 6300.3 and 6363.8 \AA, and the Na doublets at 5672$-$88 and at 6154$-$60 \AA. 
Both \citet{ivans01} and \citet{lai02} used 
the same lines to derive Na abundances, but \citet{ivans01} determined the O abundance from the O triplet lines
at 7770 \AA. All studies carried out the usual NLTE corrections from \citet{gra05}.

Na and O abundances for a large number of stars in M13 were extensively studied by \citet{johnson01} 
and \citet{sneden01}. \citet{johnson01} did not apply corrections
for non-LTE effects, while \citet{sneden01} applied the correction using the suggested procedure
by \citet{gra05}.

We matched stars with abundance information from the literature based on their 2MASS coordinates 
with the RA, DEC found in the Stetson database. Because APOGEE is only able to observe the brightest stars 
we limited the search to V$<$16. For the majority of our targets the match was easily achievable, but we 
did not include those stars in our analysis for which the positional differences were larger than 1 arcsecond. This 
resulted only in a handful of rejections and their exclusion has a minimal impact on our science results. 
The GC M2 is the only Gaia-ESO cluster \citep{pancino01} in common with APOGEE. Table 1 lists all 
(even the ones not discussed in detail) 
literature sources that were used to collect abundances of Na and O, while abundances of 
C, N, Mg and Al were only taken from \citet{meszaros01}. The photometric magnitudes of stars and their
abundances of C, N, Mg and Al are listed in Table 2, while other literature abundances of Na 
and O are listed in Table 3.

\begin{deluxetable*}{llrrcl}[!ht]
\tabletypesize{\scriptsize}
\tablecaption{Properties of the Studied Clusters}
\tablewidth{0pt}
\tablehead{
\colhead{ID} & \colhead{Name} & \colhead{N\tablenotemark{a}} &
\colhead{[Fe/H]\tablenotemark{b}} & \colhead{Literature\tablenotemark{c}} & 
\colhead{Shape\tablenotemark{d}} }
\startdata
NGC 7078	& M15		& 23 & -2.28 &  a, b, c, d	& bimodal/continuous \\
NGC 6241 	& M92		& 47 & -2.23 &	n, o		& bimodal/continuous \\
NGC 5024	& M53		& 15 & -1.95 & 	d			& bimodal\\
NGC 6205	& M13		& 81 & -1.50 &  d, e, f, g, h	& continuous 	\\
NGC 7089	& M2		& 18 & -1.49 &   d			& bimodal/continuous \\
NGC 5272	& M3		& 55 & -1.40 & 	d, i, j, k, l	& bimodal\\
NGC 5904	& M5		& 121 & -1.24 &  a, d, j, k, l, m 	& continuous \\
\enddata
\tablenotetext{a}{N is the number of stars analyzed in this paper.}
\tablenotetext{b}{[Fe/H] reference: \citet{meszaros01}.}
\tablenotetext{c}{Literature abundances: (a) \citet{carretta03}, (b) \citet{sneden05}, 
(c) \citet{sobeck01}, (d) \citet{meszaros01}, (e) \citet{johnson01}, (f) \citet{cohen01}, 
(g) \citet{kraft02}, (h) \citet{sneden01}, (i) \citet{cavallo01}, (j) \citet{sneden03}, 
(k) \citet{ivans01}, (l) \citet{lai02}, (m) \citet{ram01}, (n) \citet{sneden02}, (o) \citet{roederer01}  }
\tablenotetext{d}{The shape of Mg-Al anticorrelation from \citet{meszaros01}.}
%\tablecomments{}
\end{deluxetable*}

\begin{deluxetable*}{lrrrrrrrrrrrrrrrrrr}[!ht]
\tabletypesize{\scriptsize}
\tablecaption{Ground-Based Photometry and abundances from APOGEE}
\tablewidth{0pt}
\tablehead{
\colhead{2MASS ID} & \colhead{Cluster ID} & \colhead{Phot. ID} & 
\colhead{U} & \colhead{B} &
\colhead{V} & \colhead{R} & \colhead{I} & 
\colhead{T$_{\rm eff}$} & \colhead{log g} & 
\colhead{[Fe/H]} & \colhead{[C/Fe]} & \colhead{[N/Fe]} & \colhead{[O/Fe]} &
\colhead{[Mg/Fe]} & \colhead{[Al/Fe]} & \colhead{[Si/Fe]} & \colhead{[Ca/Fe]} & \colhead{[Ti/Fe]}
}
\startdata
2M21301565+1208229 & M15 & 73829 & 15.492 & 15.123 & 14.116 & 99.999 & 12.919 & 4836 & 1.56 & -2.12 & 9999 & 9999 & 9999 & 0.16  & -0.06 & 0.35 & 0.19 & 9999 \\
2M21301606+1213342 & M15 & 74154 & 15.635 & 15.378 & 14.408 & 99.999 & 13.239 & 4870 & 1.64 & -2.31 & 9999 & 9999 & 9999 & 0.1   & 0.57  & 0.46 & 0.53 & 9999 \\
2M21304412+1211226 & M15 & 85742 & 15.384 & 14.765 & 13.635 & 99.999 & 12.343 & 4715 & 1.28 & -2.12 & 9999 & 9999 & 9999 & -0.45 & 0.63  & 0.6  & 0.35 & 9999 \\
2M21290843+1209118 & M15 & 5875  & 15.188 & 14.599 & 13.504 & 99.999 & 12.226 & 4607 & 1.03 & -2.07 & 9999 & 9999 & 9999 & -0.11 & 0.75  & 0.41 & 9999 & 9999 \\
2M21294979+1211058 & M15 & 28871 & 15.246 & 14.338 & 13.098 & 99.999 & 11.648 & 4375 & 0.56 & -2.31 & -0.44 & 0.95 & 0.44 & 0.17 & 0.64  & 0.44 & 0.06 & 9999 \\
\enddata
\tablecomments{This table is available in its entirety in machine-readable form in the online journal. A portion 
is shown here for guidance regarding its form and content. 
Photometry is from the collection of Peter Stetson, the abundances are from \citet{meszaros01}.}
\end{deluxetable*}

\begin{deluxetable*}{lrrrrrrrrc}[!ht]
\tabletypesize{\scriptsize}
\tablecaption{Photometry and Na and O abundances from the literature}
\tablewidth{0pt}
\tablehead{
\colhead{2MASS ID} & \colhead{U} & \colhead{B} &
\colhead{V} & \colhead{R} & \colhead{I} & 
\colhead{[Fe/H]} & \colhead{[O/Fe]} & \colhead{[Na/Fe]} &
\colhead{Literature } }
\startdata
2M21295311+1212310 & 15.276 & 14.871 & 13.826 & 99.999 & 12.611 & -2.306 & 0.323 & 0.204 & a \\
2M21295492+1213225 & 15.032 & 14.166 & 12.863 & 99.999 & 11.433 & -2.225 & 0.540 & -0.072 & a \\
2M21294359+1215473 & 15.520 & 15.282 & 14.313 & 99.999 & 13.157 & -2.335 & 0.278 & 0.041	& a \\
2M21291235+1210498 & 15.293 & 14.694 & 13.567 & 99.999 & 12.283 & -2.303 & -0.092 & 0.703 & a \\
2M21294693+1208265 & 15.877 & 15.751 & 14.892 & 99.999 & 13.758 & -2.351 & 0.660 & 0.003 & a \\
\enddata
\tablecomments{This table is available in its entirety in machine-readable form in the online journal. A portion 
is shown here for guidance regarding its form and content. 
Photometry is from the collection of Peter Stetson, the abundance literature sources are listed in Table 1.}
\end{deluxetable*}

\section{Second Generation AGB stars in M92}

\begin{deluxetable*}{lrrrrrrrrrrrrr}[!ht]
\tabletypesize{\scriptsize}
\tablecaption{Photometry and abundances information of AGBs in M92}
\tablewidth{0pt}
\tablehead{
\colhead{2MASS ID} & \colhead{U} & \colhead{B} &
\colhead{V} & \colhead{R} & \colhead{I} & 
\colhead{T$_{\rm eff}$} & \colhead{log g} & 
\colhead{[Fe/H]} & \colhead{[C/Fe]} & \colhead{[N/Fe]} & \colhead{[O/Fe]} &
\colhead{[Mg/Fe]} & \colhead{[Al/Fe]}}
\startdata
\cutinhead{First generation stars}
2M17165738+4307236 & 14.283 & 13.624 & 12.479 & 11.841 & 11.189 & 4518 & 0.84 & -2.17 & -0.26 & 0.90 & 0.68 & 0.37 & -0.19  \\
2M17171342+4308305 & 14.236 & 13.574 & 12.419 & 11.779 & 11.124 & 4504 & 0.80 & -2.23 & -0.51 & 0.51 & 0.66 & 0.42 & -0.12  \\
2M17171043+4311076 & 14.260 & 13.544 & 12.357 & 11.699 & 11.039 & 4410 & 0.57 & -2.30 & -0.57 & 1.09 & 0.67 & 0.33 & -0.20  \\
2M17163772+4308411 & 14.685 & 14.532 & 13.723 & 13.233 & 12.716 & 4974 & 1.87 & -2.17 & \nodata & \nodata & \nodata & 0.16 & -0.37  \\
2M17171654+4310449 & 14.405 & 14.058 & 13.073 & 12.509 & 11.918 & 4648 & 1.14 & -2.38 & \nodata & \nodata & \nodata & 0.41 & -0.29  \\
\cutinhead{Second generation stars}
2M17170588+4310171 & 14.429 & 14.051 & 13.100 & 12.548 & 11.962 & 4729 & 1.31 & -2.26 & \nodata & \nodata & \nodata & 0.25 & 0.68  \\
2M17170033+4311478 & 14.796 & 14.667 & 13.870 & 13.382 & 12.868 & 5007 & 1.95 & -2.35 & \nodata & \nodata & \nodata & 0.34 & 0.83  \\
2M17170538+4309100 & 14.799 & 14.726 & 13.974 & 13.503 & 13.019 & 4830 & 1.53 & -2.38 & \nodata & \nodata & \nodata & 0.06 & 0.73  \\
2M17163427+4307363 & 14.680 & 14.504 & 13.677 & 13.166 & 12.659 & 4864 & 1.61 & -2.10 & \nodata & \nodata & \nodata & 0.10 & 0.25  \\
2M17172157+4307408 & 14.550 & 14.293 & 13.443 & 12.926 & 12.364 & 4868 & 1.61 & -2.37 & \nodata & \nodata & \nodata & 0.21 & 1.11  \\
\enddata

\end{deluxetable*}

As mentioned in the introduction, SG-AGB stars had been found in many GCs, except for the most metal 
poor clusters, those below [Fe/H]$<-$2. 
Here, we use the same technique first employed by \citet{garcia03}, and report on the discovery of 
SG-AGB stars in one of the most metal-poor GCs, M92, extending the covered metallicity range of observed 
GCs with SG-AGB stars down to [Fe/H]$=-$2.23.

By combining ground based photometry with Al abundances from \citet{meszaros01}, we were able to expand 
the sample of SG-AGB stars by identifying 10 AGB stars in M92, five of them first generation (FG) and five 
second generation (SG). 
Figure 1 shows three different CMDs of M92: $U-(U-I)$, $I-(U-I)$ and $V-(B-I)$. AGB stars generally 
separate most from the RGB stars in the $U-(U-I)$ CMD, and we used this CMD to identify AGB stars. 
Table 4 lists the AGB stars sampled in M92. Additional validation of these stars being SG ones could be done
by examining their O abundances (as all SG stars are O poor); all of our SG-AGB
stars are, however, hotter than 4500K, which made it impossible to measure their
[O/Fe] values.

With the discovery of SG-AGB stars in M92, the number of clusters with evidence of multiple populations 
along the AGB rises to 6. Since M92 is one of the most metal poor clusters in the Galaxy, the presence of SG-AGB stars 
in this GC provides sound evidence of the fact that these stars are commonly present in all GCs regardless 
of the cluster properties such as its mass and metallicity. As pointed out by \citet{garcia03}, the lack 
of previous evidence for SG-AGB stars was - at least partially - due to the use of less precise optical-band photometry, that 
did not allow a reliable separation of AGB and RGB stars. Another possibility is that non-LTE effects in AGB stars 
are larger than in RGB stars, which results in higher Na abundances, so only using Na to separate SG-AGB stars 
from FG stars may be misleading. The adoption of Al abundances from the APOGEE 
survey may circumvent this problem, although the effect of NLTE on Al lines in the H-band is still under 
investigation \citep{nord01} and will become available in future APOGEE data releases. 
Nevertheless, the combination of multiple abundances known 
to vary between multiple populations (Na, Al, N) with photometric data is the most accurate way to identify SG-AGB stars.

\section{Results Based on Ground Based Photometry}

\begin{figure*}[!ht]
\centering
\includegraphics[width=4.4in,angle=270]{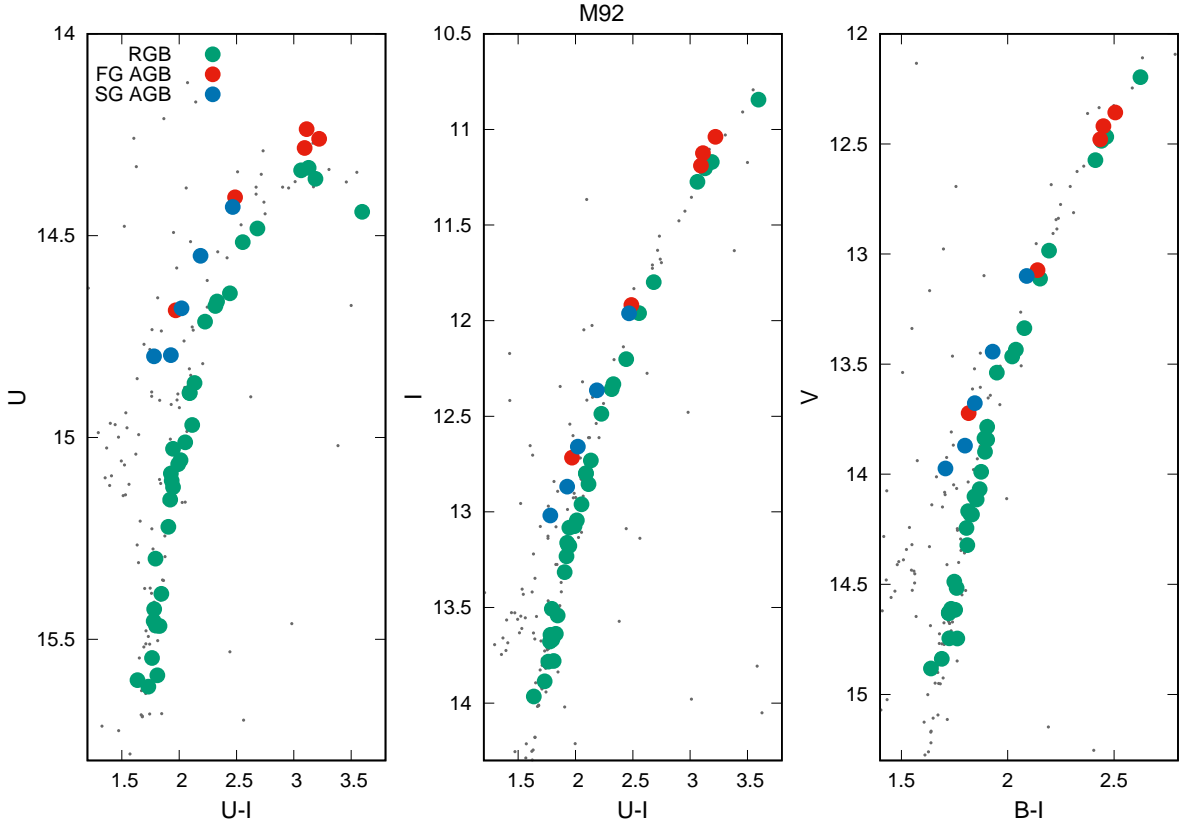}
\caption{First (red dots) and second (blue dots) generation AGB stars identified by their positions 
in the $U-(U-I)$, $I-(U-I)$ and $V-(B-I)$ diagrams. Regular RGB stars that have spectroscopic information 
from \citet{meszaros01} are denoted by green circles. 
}
\label{fig:phot1}
\end{figure*}

\begin{figure*}[!ht]
\centering
\includegraphics[width=4.4in,angle=270]{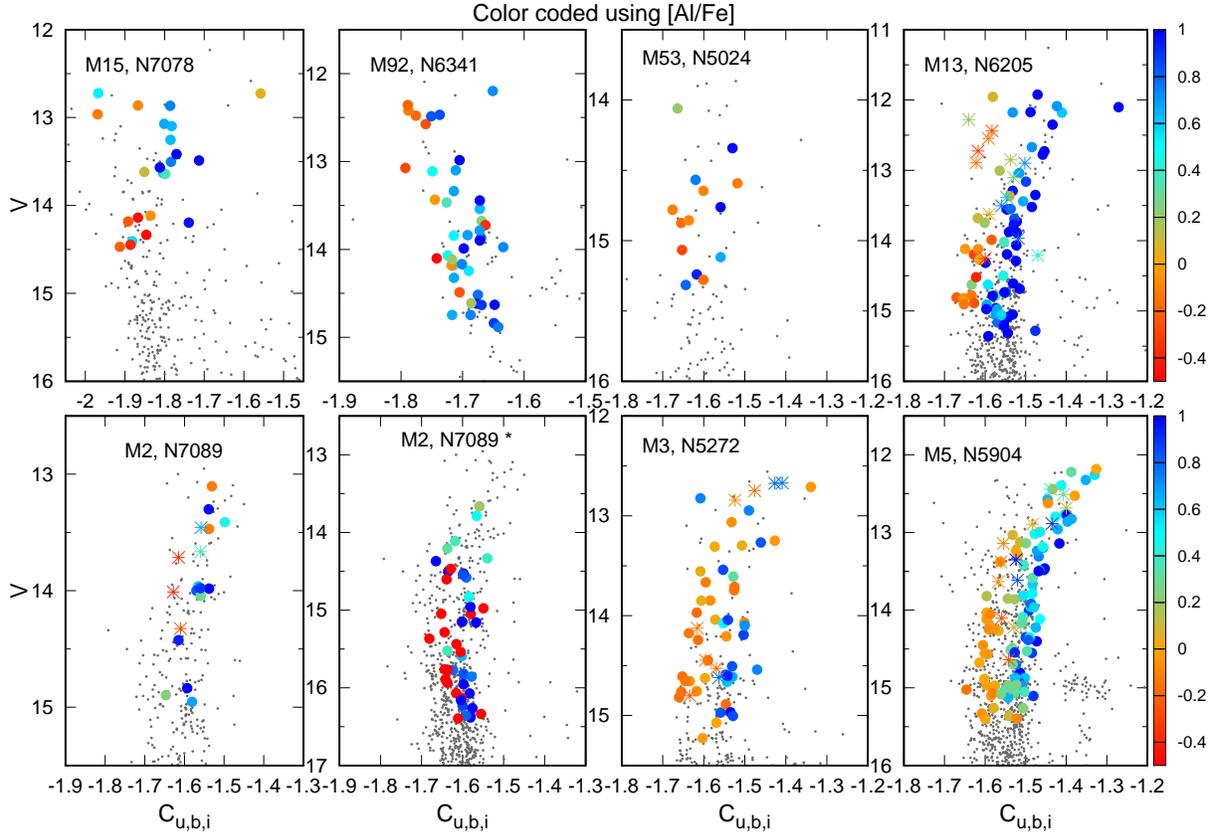}
\caption{V-Cubi diagram colour coded by [Al/Fe]. AGB stars are denoted by stars, RGB stars are by circles.
The second plot of M2 marked by a star shows data from the Gaia-ESO survey \citep{pancino01}. 
The more Al-rich SG stars occupy the left side of the RGB, while the more Al-poor FG stars are on the right
side of the RGB.
}
\label{fig:phot2}
\end{figure*}

\begin{figure*}[!ht]
\centering
\includegraphics[width=4.4in,angle=270]{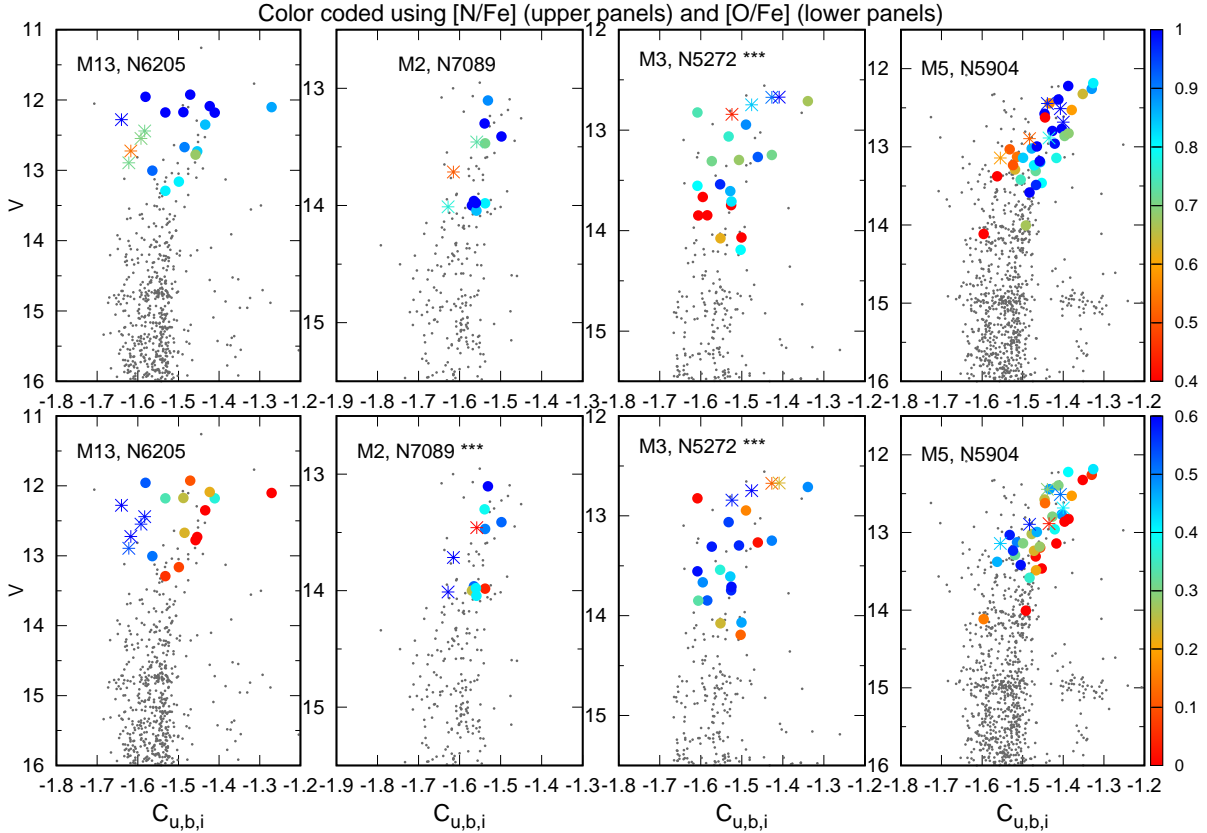}
\caption{V-Cubi diagram colour coded by [N/Fe] and [O/Fe]. 
AGB stars are denoted by stars, RGB stars are by circles. FG and SG stars divide the RGB similarly to what can be seen in 
Figure 2. Clusters with $[M/H]<-1.8$ are not plotted, because the uncertainties of [N/Fe] and [O/Fe] 
are high \citep{meszaros01}. 
}
\label{fig:phot4}
\end{figure*}

\begin{figure*}[!ht]
\centering
\includegraphics[width=4.5in,angle=0]{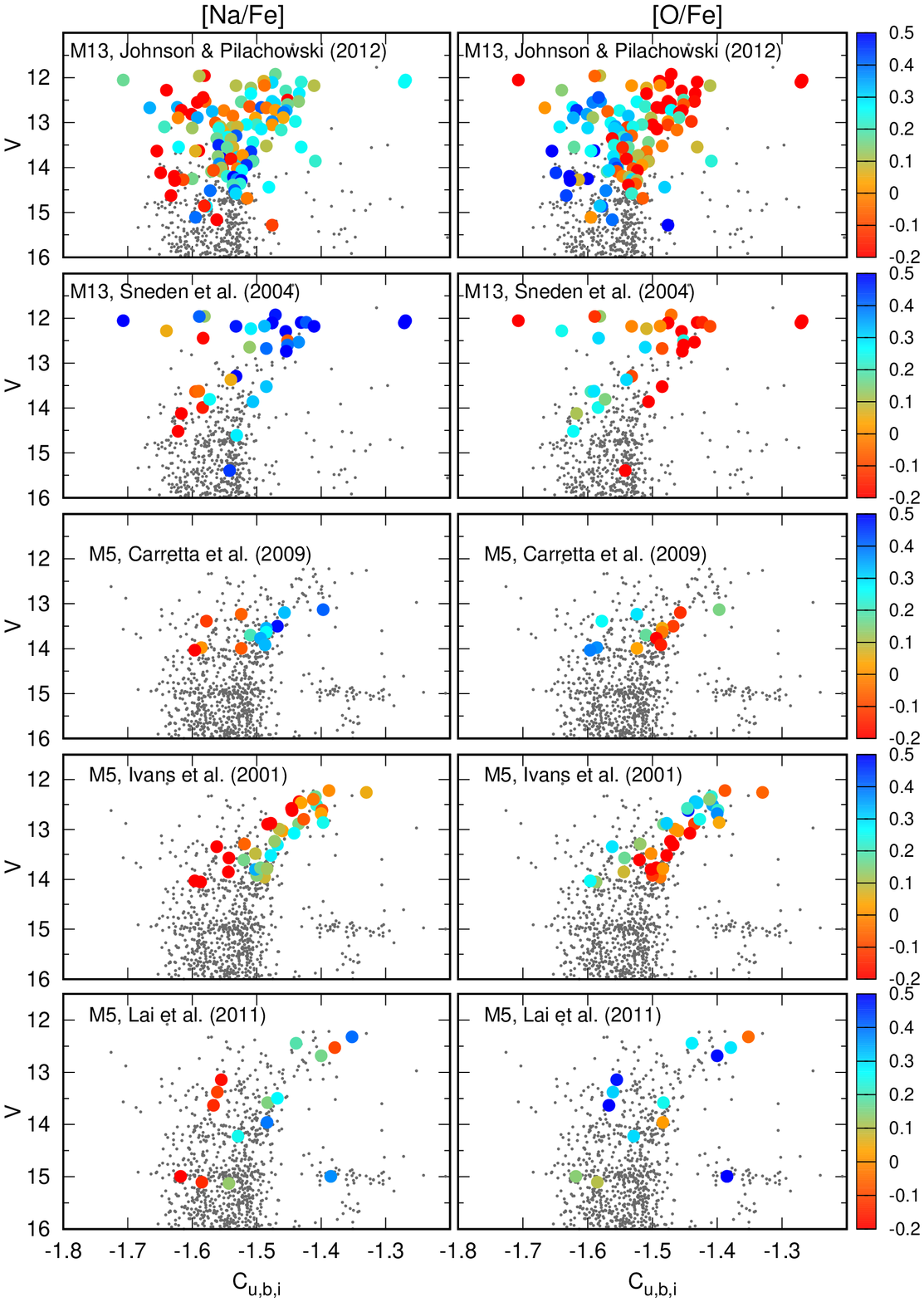}
\caption{V-Cubi diagram colour coding using literature data of [Na/Fe] and [O/Fe]. FG and SG stars are again well separated 
on the RGB for most clusters.
}
\label{fig:phot6}
\end{figure*}

The first detailed theoretical investigation of the impact of the peculiar chemical patterns of multiple 
stellar populations on the stellar spectral energy distribution was performed by \citet{sbordone01}. They 
found that CNO element abundance variations do affect the stellar spectra essentially at wavelengths shorter 
than about 400 nm; i.e. the UV spectral windows. This evidence provided a plain support to the use of UV 
photometric passbands - and combination of UV and optical bands - to properly trace the presence/properties 
of multiple stellar populations in GCs. Later, \citet{cassisi03} - by using synthetic spectra computed 
for appropriate light element distributions - showed that the 
Mg-Al anti-correlation has no impact on the stellar models and isochrones, as opposite to C-N and O-Na.

The \cubi $= (U - B) - (B - I)$ photometric index was first introduced by \citet{milone04} in 47 Tuc. They observed   
that a pseudo-CMD based on this index is very sensitive to the composition of stars making the different 
populations stand out in the \cubi$-$V diagram. This is the result of the \cubi index being affected by 
changes in the UV flux caused by variations in [N/Fe]. Elements that correlate with N, like Na and Al
can also be used with the \cubi index to identify FG and SG stars, even though none of them has a direct effect on the UV flux.

As \citet{meszaros01} noted, some northern clusters show bimodal Al distributions (M53, M3), while other clusters exhibit continuous 
distribution (M5, M13). At the same time we can clearly observe bimodal and/or more
continuous RGB branches in both ground-based \cubi and HST
\chst colour indexes (see next Section). Because these colour
indexes are not directly sensitive to Al variations, we can only explore relationship between
abundances and photometry indirectly with the
shape of the Mg-Al anticorrelation and the N induced photometric variations.

Previously, \citet{monelli01} was able to systematically study the 
behavior of the \cubi index in 15 clusters, and they showed that O abundances clearly correlated with 
the \cubi index in most cases. This made it possible to use only ground-based photometry to
easily separate first and second population stars in globular clusters as they are split into two distinct 
groups in the \cubi pseudo-CMD. This technique was also used by \citet{lardo02} in M4, 
\citet{milone02, milone03} in M2, NGC~2808, and \citet{nardiello01} in NGC 6752, NGC 6397, and M4. 
Other indexes can also be used, like $(U - V) - (V - I)$, but none of them are as sensitive to variations 
of CN molecular bands in the optical as the \cubi.

\citet{monelli01} did not use Al in their study, but because [Na/Fe] and [Al/Fe] correlate with [N/Fe] in the 
metal-poor clusters below [Fe/H]=$-$1 \citep[see, e.g.,][and references therein]{meszaros01}, we expect to see a clear 
separation in both Al and N. Figure 2 and 3 show the pseudo-CMD of all seven clusters; stars with known Al, N and O 
abundances are coloured according to their abundance values. In the traditional CMDs, we cannot see the 
split RGB belonging to first and second populations. However, in the \cubi pseudo-CMD, the separation is clearly 
highlighted when a different colour coding based on the measured Al abundance is adopted. 
First generation stars with low [Al/Fe] content have generally lower \cubi index, 
as can be seen in both the APOGEE and Gaia-ESO data \citep{pancino01}.
Interestingly, the separation becomes less clear at higher luminosities; above V$<$13 more first 
generation stars are mixed with second generation stars, this is particularly noticeable in in M53, M2, M3, and M5. 
This can be explained with the behavior of the \cubi index. The two main RGB branches have very similar
\cubi values at high luminosities because the part of the UV and optical spectra where the CN bands can be found 
loses its sensitivity to the variation of the N abundance. By overplotting 
the AGB stars on the \cubi $-$ V diagram (Figure 2) we find that AGB stars are not well separated from the RGB stars.

Besides studying the behavior of Al in the \cubi $-$ V diagram we are also able to plot the literature 
[Na/Fe] and [O/Fe] values used (Figure 4). As expected from previous studies and from the Al-O and Na-O anticorrelations, 
both elements can be used to separate first and second generation stars, as reported previously. 
In M13, as shown in Figure 4, the various sub-populations are not well separated when using 
the Al abundances from \citet{johnson01} to trace them. We note that \citet{johnson01} did not account 
for any non-LTE effects when estimating their [Na/Fe] abundances. 
However, the non-LTE correction should amount to 0.1~dex at most \citep{gra05}. 
So the most plausible explanation for this result is likely due to the 
lower quality (moderate resolution combined with an extremely short wavelength coverage) 
spectral data used by \citet{johnson01}. This
combined with the fact that the RGB sequences converge towards the RGB tip
likely explain the apparent problem.

From data in Figures 2 to 4 it appears evident that the separation along the RGB between FG and SG stars 
is significant, but not perfect with the most blended cases being those corresponding to M2, M3 and M53. In M92 there is one 
star with low Al in the SG branch. In M53 there are three stars with high Al values in the FG branch, and in M3
there are at least five stars with low Al in the SG RGB branch. In M2, the Gaia-ESO data shows three low Al 
stars in the SG branch, while the separation is clearer in the APOGEE data; however APOGEE sampled fewer and 
more luminous stars. We have to note that besides these small differences, both surveys agree very well in 
terms of identifying FG and SG stars. At present, the explanation of
these apparent outlier stars is not clear. Possible reasons are: i) random errors in the
data reduction and/or in the Al abundances spectroscopic determination; and ii)
errors in the photometry. 
The latter is more probable because the U-band magnitudes have generally higher errors than other filters, while both 
APOGEE and Gaia-ESO use high-resolution, high S/N spectra and errors up to 1~dex in [Al/Fe] are very unlikely. 
Finally, it is also possible that the explanation
lies in an astrophysical origin, but in order to prove that, a careful
examination of all the above possible errors would be necessary, which is beyond
the scope of this paper.

\section{HST Photometry in Combination with Mg-Al}

\begin{figure*}[!ht]
\centering
\includegraphics[width=6.5in,angle=0]{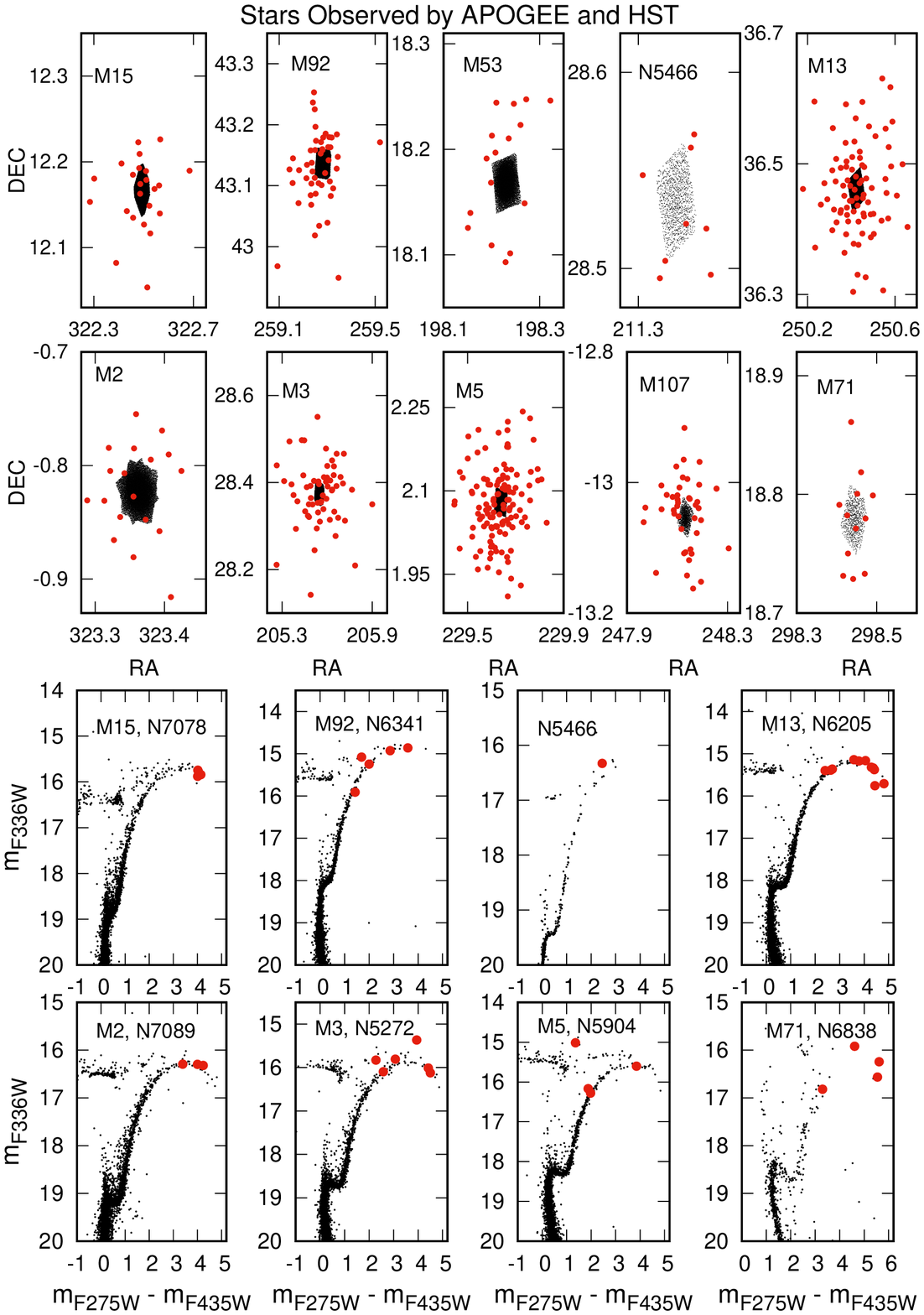}
\caption{HST CMDs (black dots) with stars that are in common with stars observed by the APOGEE survey (red dots). 
HST observes only the centre of each cluster, so the overlap is generally small. 
}
\label{fig:hst}
\end{figure*}

\begin{figure*}[!ht]
\centering
\includegraphics[width=4.4in,angle=270]{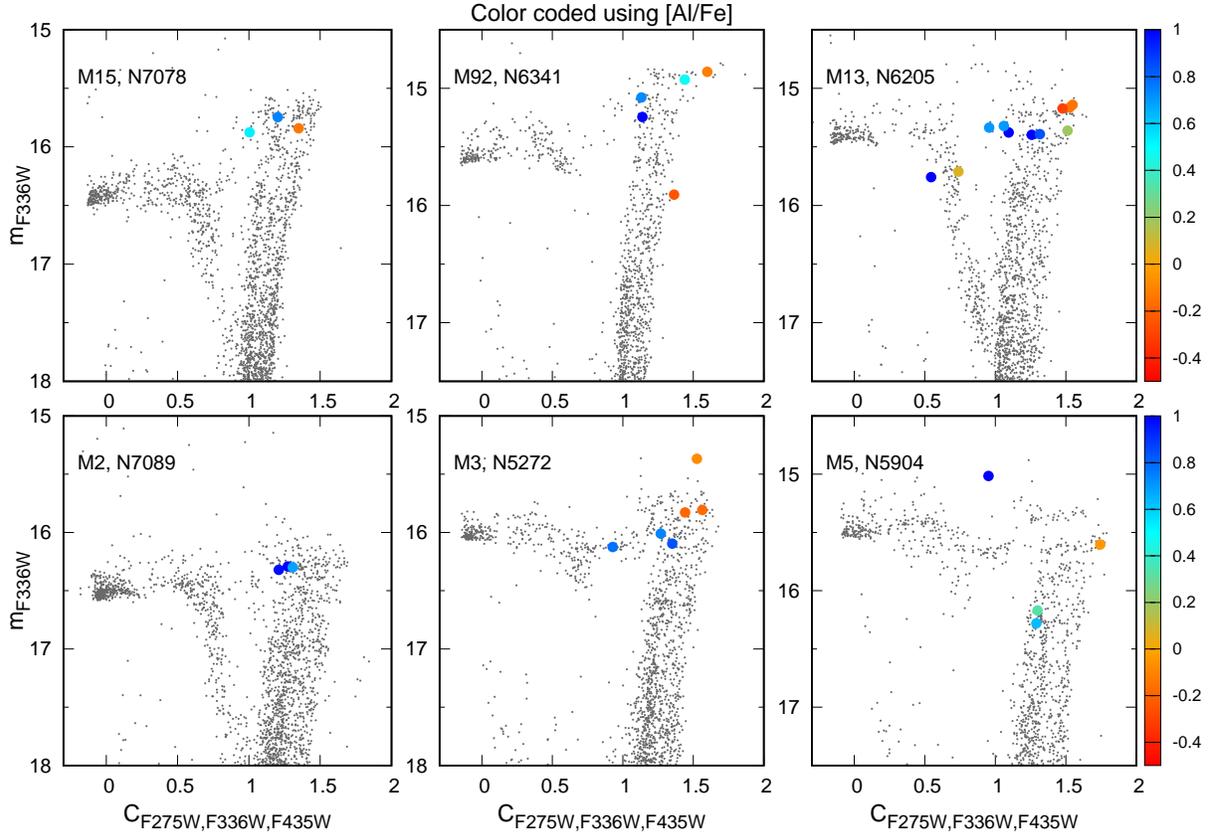}
\caption{HST pseudo-CMDs overplotted by stars in common with APOGEE colour coded by their [Al/Fe]. 
SG Stars with high Al content are on the left branch of the RGB, and FG stars are on the right.
}
\label{fig:hst2}
\end{figure*}

\begin{figure*}[!ht]
\centering
\includegraphics[width=5.5in,angle=0]{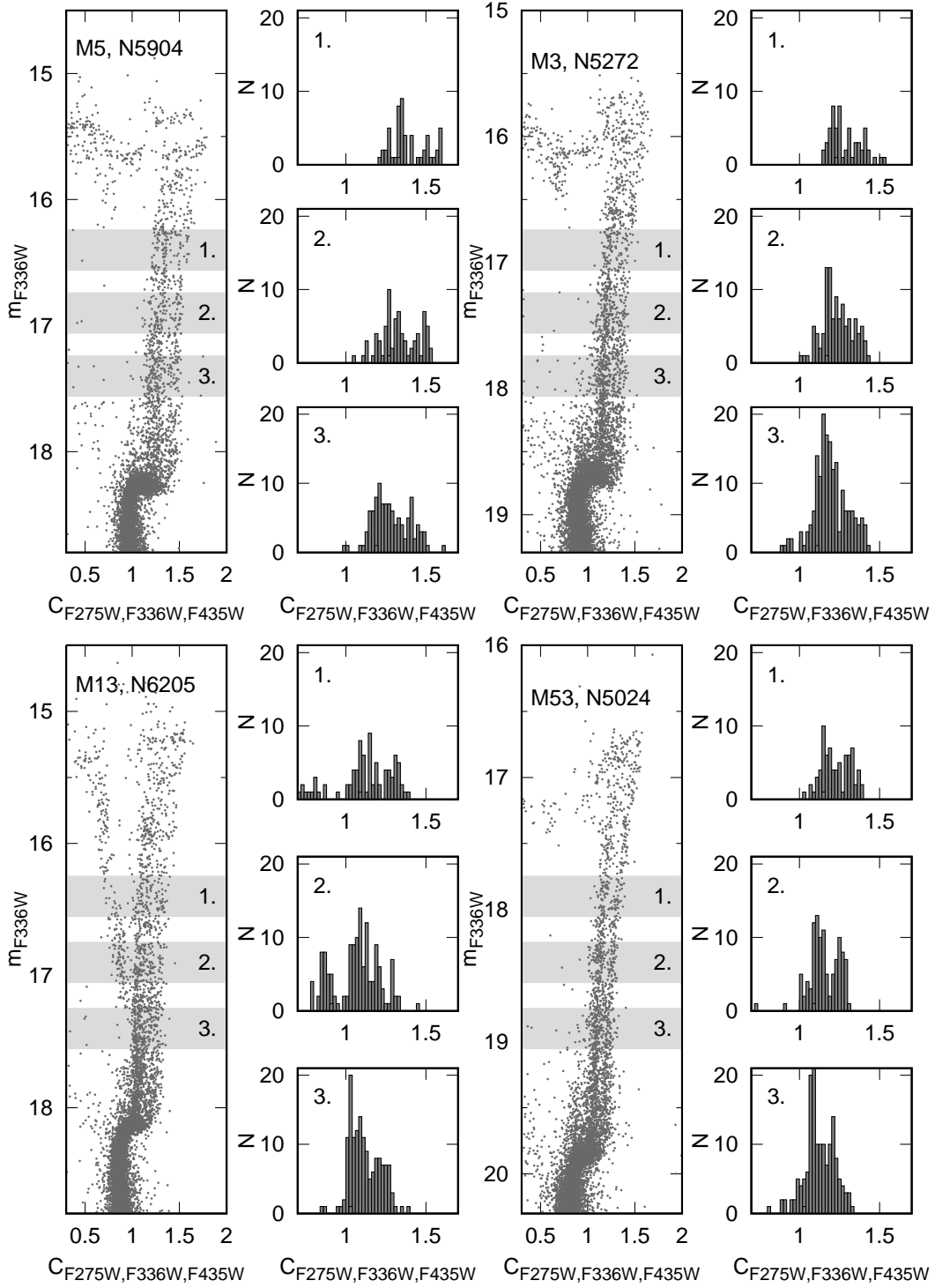}
\caption{The histograms show the number of stars found in three 0.5 magnitude wide regions that 
are 0.5 magnitude far from each other starting 1.5 magnitude down from the tip of the RGB. Clusters
with bimodal Al distribution are M3 and M53, continuous distribution are M13 and M5. There is no visible connection
between the shape of the Mg-Al anticorrelation and the histogram of the number of stars found in the RGB. 
}
\label{fig:hst3}
\end{figure*}

\begin{deluxetable*}{lrrrrrrrrrrr}[!ht]
\tabletypesize{\scriptsize}
\tablecaption{HST Photometry and abundances from APOGEE}
\tablewidth{0pt}
\tablehead{
\colhead{2MASS ID} & \colhead{Cluster} & \colhead{m$_{F275W}$} &
\colhead{m$_{F336W}$} & \colhead{m$_{F435W}$} & \colhead{HST ID} &
\colhead{[Fe/H]} & \colhead{[C/Fe]} & \colhead{[N/Fe]} & \colhead{[O/Fe]} &
\colhead{[Mg/Fe]} & \colhead{[Al/Fe]}}
\startdata
2M21295678+1210269 & M15 & 18.381 & 15.876 & 14.376 & 184919 & -2.31 & -0.36 & 1.36 & 0.66 & 0.34 & 0.53  \\
2M21300274+1210438 & M15 & 18.353 & 15.746 & 14.341 & 141485 & -2.32 & \nodata  & \nodata & \nodata & 0.31 & 0.74  \\
2M21295666+1209463 & M15 & 18.586 & 15.842 & 14.447 & 104619 & -2.27 & -0.54 & 1.25 & 0.62 & 0.30 & -0.14 \\
2M17170731+4309308 & M92 & 17.075 & 14.926 & 14.216 & 127737 & -2.10 & \nodata  & \nodata & \nodata & 0.19 & 0.48  \\
2M17171342+4308305 & M92 & 17.451 & 14.860 & 13.868 & 76921  & -2.23 & -0.51 & 0.51 & 0.66 & 0.42 & -0.12 \\
\enddata
\tablecomments{This table is available in its entirety in machine-readable form in the online journal. A portion 
is shown here for guidance regarding its form and content. 
Photometry is from the HST Treasury Project \citet{soto01, piotto02}, the abundances are from \citet{meszaros01}.}
\end{deluxetable*}

\subsection{Observed Properties}

Using the preliminary data release of the HST Treasury Project \citep{soto01, piotto02} 
we were able to match stars 
from \citet{meszaros01} with their HST catalog 
using 2MASS coordinates and a magnitude cut of 18 or 19 in m$_{F336W}$, depending on clusters 
to avoid contamination from fainter stars. 
The field of view of 
the HST is small compared to that of 2.5-meter SDSS telescope \citep{gunn01} and the HST observations focused on the 
centre of each cluster, while APOGEE observed mostly their outer parts, this resulted in relatively 
few matches. Altogether, we found 36 stars 
(four of them AGB stars) in 8 clusters in common between APOGEE and HST.
As can be seen from Figure 5, the few common stars are usually among the brightest 
ones near the tip of the RGB, but this still allows us to discriminate between 
the multiple branches of the RGB due to the high quality of the HST photometry and associate these 
different RGB with a chemical
composition.

This can be done using the $m_{\rm F336W}$ $-$ \chst pseudo-CMD displayed in Figure 6. This index
is a similar diagnostic tool to \cubi in that it is also sensitive to the [N/Fe] content of stars and separates multiple
populations from each other well. As previously mentioned, one can only use the \cubi index to indirectly associate 
to an Al abundance since the \cubi is sensitive to N and not Al directly, and this is also true for the \chst index.

By overplotting the stars with [Al/Fe] abundances from the APOGEE survey 
we can conclude from Figure 6 
that Al-rich stars are well separated from the Al-poor ones at the top of the RGB; the SG RGB stars have 
slightly lower \chst index than FG stars. This is very similar to the behavior we see when using the \cubi index in Figure 2. 
From the \cubi index we know that this separation continues down to the turn-off, as shown for 47 Tuc by \citet{milone05} and 
for NGC~2808 by \citet{milone03}. We expect the same to be true for the \chst index as well, even though we do not have 
[Al/Fe] available for such low luminosity stars.

The extremely precise photometry of HST allows us to examine any possible correlation in the structure of the RGB, and 
the shape of the Mg-Al anti-correlation for all 10 GCs in \citet{meszaros01}. 
Stellar members from \citet{meszaros01} in common with the HST photometry are listed in Table 1. The additional three GCs
are NGC 5466 ([Fe/H]=$-$1.82), M107 ([Fe/H]=$-$1.01), and M71 ([Fe/H]=$-$0.68). NGC~5466 
displays a bimodal Mg-Al anticorrelation, although only 8 stars were
observed by APOGEE, and only two were SG. The GCs M107 and M71, however, do not display a Mg-Al
anticorrelation, as expected from their high metallicities (see below). 

In order to investigate possible connection between the photometric and abundance distributions, 
we contrast the shape of the Mg-Al anticorrelation 
against the histogram of the number of stars found in the RGB using HST magnitudes in M5, M13, M3 and M53 (Figure 7). 
M5 and M13 are clear examples of continuous Al distributions, while M3 and M53
are clear examples of bimodal distributions of Al. Other clusters, such as M15, M92,
and M2 fall somewhere in between and in order to detect any possible correlation, we
chose to show examples of the most extreme distributions. We do this by defining three 0.5 magnitude wide regions that 
are separated by 0.5 magnitudes starting 1.5 magnitude down from the tip of the RGB and create a histogram
of the number of stars found in each of these regions. We conclude from Figure 7 that there are no visible connection
between the shape of the Mg-Al anticorrelation and the histogram of the number of stars found in the RGB. M3 and M53 
have very clear distinctive Al-rich and Al-poor populations, while the distribution of the RGB branches are no more 
distinctive than those of M5 and M13. In fact, M5, which has a continuous Mg-Al distribution, has two distinct 
peaks in the histogram, while M3 is the opposite. We believe this either rules out a GC formation scenario with two separate 
star forming events with no new stars forming in between them, or the time spent between them was not enough the push
the two branches of RGBs so far from each other to be visible. Alternatively, the multiple star formation burst are overlapping. 
From Figure 7, one can also confirm that the Al content has no effect on the structure of the \chst pseudo-CMD, 
agreeing with the theoretical understanding of \citet{cassisi03}.

\subsection{Possible Interpretation of Observed Properties}

In the last decade, several scenarios have been suggested to explain the origin of multiple stellar populations 
in Galactic GCs, including: fast rotating massive stars \citep{decressin01}, 
interacting massive binary stars \citep{demink01}, accretion on circumstellar disk during the Pre-MS 
stage \citep{bastian01, cassisi04}, supermassive MS stars \citep{deni01}, 
and massive AGB stars \citep{ventura01, dercole02}. 
Each one of the proposed scenario can reproduce some observational evidence, 
no one of them is able to provide a plain interpretation of the observational framework. 
All of them have their specific pro and cons \citep[see][for a detailed discussion on this issue]{renzini01}.
Even though we are well aware of strong limitation that all mentioned scenarios have, 
since in this work we focus on the Mg-Al anticorrelation, here we rely on the 
AGB scenario that (to the best of our knowledge) has been so far the unique one 
able to provide useful hints on the Al distribution observed in metal-poor ([Fe/H]$< -$1) GGCs \citep{ventura03}. 
More recently, \citet{dell01} have successfully modeled 
the Mg-Al anti-correlation in nine GCs observed by APOGEE ($-2.2<$[Fe/H]$<-0.7$) and found
remarkable agreement between the observations and theoretical yields from 
massive AGB stars, supporting the earlier \citet{ventura03} results on a 
smaller APOGEE GCs sample. This further 
supports the idea that the main driving force of pollution is the ejecta of AGB stars 
in the range of metallicities considered. For these reasons we concentrate this discussion on the AGB scenario.

From Figure 7 we found that in M3 and M53 the separation of multiple RGB branches in the \chst diagram does 
not correlate with the discreetness of the Mg-Al anticorrelation. The difference between discrete and 
continuous Mg-Al distributions could be explained by the dilution of AGB ejecta with pristine
gas; \citet{ventura03} found that the Mg-Al anticorrelation can be explained
by theoretical yields from massive AGB stars with different dilution levels. 
According to \citet{ventura03} SG stars in M3 formed from gas that contained at least 30\% diluted
material from FG stars, while there were no stars that formed from 10$-$30\%
diluted gas. The situation is different in M13 and M5, two clusters with
continuous distribution of Al abundances, where stars formed from all fractions
of dilution, from 0 to 100\%. Under the AGB self-enrichment hypothesis, the
timing of the return of pristine gas in the central regions of the cluster,
after the end of SNe II explosions, is the key factor determining the shape
(bimodal vs. continuous) of the Mg-Al anticorrelation \citep{dercole01}. 

If the duration of the process is longer than 40$-$50Myrs, a portion of SG stars 
form from non diluted matter and we thus expect two clear distinct populations; i.e., 
a bimodal Mg-Al anticorrelation with FG and SG stars showing the initial 
chemistry and very low-Mg/high-Al, respectively. Conversely, in case of a 
prompt return of the pristine gas, we then expect a continuous Mg-Al anticorrelation, 
with different dilution degrees of the AGB ejecta with pristine gas. The main factors 
affecting the timing of the return of pristine gas are the initial density 
distribution of the cluster and the number of SNe II explosions, the latter 
being determined by the total mass of the cluster. One of the main findings 
emerging from the study by \citet{dercole01} is that for a given density 
distribution the return of the pristine gas takes longer for more massive clusters. 
This result, in conjunction with the high sensitivity 
of the degree of the hot bottom burning nucleosynthesis experienced by 
massive AGBs to the metallicity, provide an explanation of the correlation between the shape 
of the Mg-Al distribution and the mass and metallicity of the clusters 
found in recent studies \citep{carretta02, carretta03, carretta01, pancino01}. 
\citet{dercole01} clearly demonstrated that the mass of the 
cluster has a strong effect on the extent of the chemical pollution patterns 
of the species touched by proton-capture nucleosynthesis. 
Indeed, the AGB self-enrichment scenario is, so far, the only one that can explain the 
increasing extension of the Mg-Al anticorrelation (chemical patterns of the chemical elements 
affected by high-temperature proton capture) observed at lower
metallicities \citep{ventura03, dell01}.

If the return of pristine gas is prompt, we then expect a continuous Mg-Al anticorrelation, with different
dilution degrees of the AGB ejecta with pristine gas. The HST UV
pseudo-CMDs are not sensitive to the Al content and we only use Al
indirectly through the expected Al-N correlation. In the AGB context, in
principle, a clear Mg-Al bimodality should be accompanied by a net separation
between N-normal and N-rich stars in the HST UV pseudo-CMDs. The CN (and also
the CNO) nucleosynthesis, which produces C-poor and N-rich gas, requires lower
temperatures than the Mg-Al chain. In AGB stars, N production begins at
$\sim$30MK, whereas Mg burning demands $\sim$90$-$100M~K. This is the reason why N
production is expected at all metallicities, while the traces of Mg-Al burning
are expected only in low- and intermediate-metallicity ([Fe/H]$\leq-$1.0) GCs,
as observed \citep[see e.g,][]{meszaros01, ventura03}.

In other words, while a net separation in N between FG and SG stars does not necessarily
require a clear separation in the Mg-Al plane (as it is still possible that the
gas was exposed to CN cycling but not to Mg-Al burning; and this is metallicity
dependent), the Al-rich SG stars must have a N content much higher than their 
FG counterparts. The latter is corroborated by our Figure 6, which shows
that SG Al-rich and FG Al-poor stars lie on the left (N-rich) and right (N-poor)
photometric HST RGB branches, respectively. 

It seems clear that there is no one-to-one correspondence between the Mg-Al
anticorrelation shape and the HST photometric information such as
the appearance of the HST-pseudo CMDs, the number and/or shape of RGB branches, and
their corresponding chromosome maps, which gives information on the presence of additional FG and/or
SG subpopulations \citep{milone06}. Both GCs with bimodal (M3 and M53) or continuous
(M5 and M13) Mg-Al anticorrelations can display a broad RGB branch with many stars
in between the main FG and SG RGB branches, and/or two rather well defined main FG
and SG RGB branches (see Figure 7). This lack of spectroscopic-photometric correspondence supports previous results 
that the HST photometric information about any single cluster usually gives a more complex star formation history than the
spectroscopic one. For example, \citet{milone03} found that NGC~2808
displays three discrete groups in the Mg-Al anticorrelation but five different
populations in the HST pseudo-CMDs and their corresponding chromosome maps. The
presence (or not) of a significant number of stars between the two main branches of the RGB
is likely related to the specific formation history of the cluster. Such stars are believed to have
intermediate N abundances (as FG is N-poor and SG is N-rich), even if they are slightly Al-rich or
not. Stars showing FG-like chemical pattern but slightly enriched in N are known to be present in 
other clusters, for example in NGC~2808 \citep{milone03}. 
\citep{dantona03} identified them as late SG stars formed at $\sim$90$-$100 
Myrs from AGB material strongly diluted with the pristine gas. We note that the occurrence 
of non canonical processes - such as extra-mixing during the RGB stage - 
could also contribute to modify the expected chemical patterns. 
Finally, it is also possible that the discreetness of the Al 
distribution is independent of the self-enrichment mechanism, perhaps because the star formation 
happens in small pockets and it is interrupted by SNe II explosions that clear out most of the gas 
from the clusters \citep{bekki01}. 

Nevertheless, in order to fully understand the connection between Al and N abundances and the \cubi, and \chst 
indexes, we need N abundances more precise than those of \citet{meszaros01}. 
Unfortunately, the available APOGEE N abundance estimates are for the most luminous RGB stars, 
near the top of the RGB where the separation between the various sub-populations is not very clear. 
While more precise N measurements will be most certainly published by the APOGEE team, 
they are still not able to measure the N content of fainter (H$<$12.5; see Figure 6) stars in the multiple RGB branches
seen by the HST.

\section{Conclusions}

The combination of photometric magnitudes and chemical information is a powerful tool in understanding
the history and evolution of GCs. It was shown by \citet{sbordone01} that certain photometric indexes sensitive 
to the abundance of N can be used to study the presence of multiple populations in GCs. 
Here, we combined these two data sets (photometric and spectroscopic) for 7 clusters and examined the behavior of 
multiple population of stars in the \cubi-V diagram. We found that first and second generation stars 
are well separated from each other in this diagram, when using elements of Al and N from APOGEE
and Gaia-ESO. We also found that the separation is less clear at the top of the RGB, because multiple 
branches converge due to the fact that \cubi loses sensitivity because the molecular bands become saturated and 
insensitive to the variations in N abundance.

We have identified 10 AGB stars in M92, 5 of them being first generation and five are second generation. 
This is the most metal-poor cluster to date in which SG-AGB stars have been found. Combined with 
\citet{garcia03}, there are now enough clusters containing SG-AGB to conclude that the appearance of SG-AGB is  
common and does not depend on the cluster's main parameters such as age, luminosity, or metallicity. 

We combined the Al abundances
from the APOGEE survey with ground-based UBVRI and HST photometry and
find that clusters with bimodal and continuous Al distributions have similar
photometric properties in both data sets. We confirm that Al does not have an effect on the structure of the \cubi 
and \chst pseudo-CMDs, as previously explained by \citet{cassisi03}.
Both GCs with bimodal/discrete (M3 and M53) or continuous (M5 and M13) Mg-Al
anticorrelations can display a broad RGB branch. This suggests, under the AGB
self-enrichment framework, that the lack of medium Al stars in M3 and M53 is
probably the result of stars not forming from 10$-$30\% diluted gas by FG
stars. 

Because there is no one-to-one correspondence between the Mg-Al anticorrelation 
shape and the photometric information such as the appearance of the HST-pseudo CMDs, 
both Mg-Al and C-N abundances are needed simultaneously in order to understand the formation 
history of each cluster and the multiple
populations phenomenon. The lack of a spectroscopic-photometric correspondence suggests that 
the HST photometric information usually gives more complex star formation histories than the
spectroscopic one. Since [N/Fe] errors 
reported by \citet{meszaros01} are relatively large (0.12$-$0.32~dex)\footnote{Note that these errors 
do not include possible systematic/random effects due to the methodology employed in the chemical 
abundances derivation; the use of spectral windows vs. the entire spectrum, model atmospheres, 
linelists, etc.}, more precise CNO 
abundances, the ability to accurately model CN yields from the internally processed material, 
and more stars to improve the statistics are definitely needed. Such data 
may be provided by the on-going APOGEE-2 survey, which will almost triple the number of GC stars 
observed in the H-band. The ideal step forward is to get spectroscopic data for
the stars observed by the HST and the different FG and SG subpopulations seen in
the chromosome maps (as already suggested by \citet{milone06}) but we likely
may have to wait for next generations instruments and/or the big telescopes era
for this; because APOGEE, only focusing on bright stars above H$<$12.5mag, will not observe
the fainter stars down in the RGB that are necessary for a complete analysis.

%\clearpage

\acknowledgements{We thank the referee for his/her suggestions during the peer-review process, which 
greatly improved the clarity of the paper. We also would like to thank Paolo Ventura for his 
useful comments and discussion when preparing this paper. 

SzM has been supported by the Premium Postdoctoral 
Research Program of the Hungarian Academy of Sciences, and by the Hungarian 
NKFI Grants K-119517 of the Hungarian National Research, Development and Innovation Office. 
DAGH was funded by the Ram\'on y Cajal fellowship number RYC$-$2013$-$14182. DAGH, FDA, TM, and OZ  
acknowledge support provided by the Spanish Ministry of Economy and
Competitiveness (MINECO) under grant AYA$-$2014$-$58082-P. SC acknowledges financial support from PRIN-INAF2014.
AD has been supported by the \'UNKP-17-4 New National Excellence Program of the Ministry of Human Capacities and 
the NKFIH K-115709 grant of the Hungarian National Research, Development and Innovation Office.
AD and LSz would like to thank the City of Szombathely for support under Agreement No. 67.177-21/2016.

Funding for SDSS-III has been provided by the Alfred P. Sloan Foundation, the Participating Institutions, the 
National Science Foundation, and the U.S. Department of Energy Office of Science. The SDSS-III web site is 
http://www.sdss3.org/.

SDSS-III is managed by the Astrophysical Research Consortium for the Participating Institutions of the SDSS-III 
Collaboration including the University of Arizona, the Brazilian Participation Group, Brookhaven National Laboratory, 
University of Cambridge, Carnegie Mellon University, University of Florida, the French Participation Group, the 
German Participation Group, Harvard University, the Instituto de Astrofisica de Canarias, the 
Michigan State/Notre Dame/JINA Participation Group, Johns Hopkins University, Lawrence Berkeley National Laboratory, 
Max Planck Institute for Astrophysics, New Mexico State University, New York University, Ohio State University, 
Pennsylvania State University, University of Portsmouth, Princeton University, the Spanish Participation Group, 
University of Tokyo, University of Utah, Vanderbilt University, University of Virginia, University of Washington, 
and Yale University.

}


\thebibliography{}


\bibitem[Alves-Brito et al.(2012)]{alves01} Alves-Brito, A., Yong, D., Meléndez, J., Vásquez, S., Karakas, A. I. 2012, \aap, 540, 3

\bibitem[Bastian et al.(2013)]{bastian01} Bastian, N. et al. 2013, \mnras, 436, 2398

\bibitem[Bekki et al.(2017)]{bekki01} Bekki, K., Jerabkova, T., Kroupa, P. 2017, arXiv, 1706.06787

\bibitem[Campbell et al.(2013)]{campbell01} Campbell, S. W., D’Orazi, V., Yong, D., et al. 2013, Natur, 498, 198

\bibitem[Campbell et al. (2017)]{camplbell02} Campbell, S. W.; MacLean, B. T.; D'Orazi, et al. A\&A (in press; arXiv: 1707.02840) 

%\bibitem[Carretta (2014)]{carretta05} Carretta, E., 2014, ApJ, 795, L28


\bibitem[Carretta et al.(2005)]{carretta04} Carretta, E., Gratton, R.~G., Lucatello, S., Bragaglia, A., \& Bonifacio, P. 2005, \aap, 433, 597

\bibitem[Carretta et al.(2009a)]{carretta02} Carretta, E., Bragaglia, A., Gratton, R., \& Lucatello, S.\ 2009a, \aap, 505, 139 

\bibitem[Carretta et al.(2009b)]{carretta03} Carretta, E., Bragaglia, A., Gratton, R.~G., et al.\ 2009b, \aap, 505, 117 


%\bibitem[Carretta et al.(2012)]{carretta06} Carretta, E., Bragaglia, A., Gratton, R. G., Lucatello, S., D'Orazi, V., 2012, \apj, 750, L14

\bibitem[Carretta et al.(2009c)]{carretta01} Carretta, E., Bragaglia, A., Gratton, R., D'Orazi, V., \&  
Lucatello, S. 2009c, \aap, 508, 695


\bibitem[Cassisi et al.(2013)]{cassisi03} Cassisi, S., Mucciarelli, A., Pietrinferni, A., Salaris, M., Ferguson, J. 2013, A\&A, 554, 19


\bibitem[Cassisi \& Salaris(2014)]{cassisi04} Cassisi, S. \& Salaris, M. 2014, \aap, 563, 10

\bibitem[Cassisi et al.(2008)]{cassisi01} Cassisi, S., Salaris, M., Pietrinferni, A. et al. 2008, \apjl, 672, L115


\bibitem[Cassisi et al.(2014)]{cassisi02} Cassisi, S., Salaris, M., Pietrinferni, A., Vink, J. S., \& Monelli, M. 2014, A\&A,
571, A81


\bibitem[Cavallo \& Nagar(2000)]{cavallo01} Cavallo, R.~M., \& Nagar, N.~M. 2000, \aj, 120, 1364

\bibitem[Charbonnel(2013)]{charbonnel01} Charbonnel, C., Chantereau, W., Decressin, T., Meynet, G., \& Schaerer, D.
2013, A\&A, 557, L17

\bibitem[Cohen \& Mel{\'e}ndez(2005)]{cohen01} Cohen, J.~G., \& Mel{\'e}ndez, J. 2005, \aj, 129, 303

\bibitem[D'Antona et al.(2005)]{dantona02} D'Antona, F., Bellazzini, M., Caloi, V. et al. 2005, \apj, 631, 868

\bibitem[D'Antona et al.(2016)]{dantona03} D'Antona, F., Vesperini, E., D’Ercole, A. et al. 2016, \mnras, 458, 2122

\bibitem[Dell'Agli et al.(2017), submitted]{dell01} Dell'Agli et al. 2017, submitted to \mnras

\bibitem[Denissenkov \& Hartwick(2014)]{deni01} Denissenkov, P., \& Hartwick, F. D. A. 2014, \mnras, 437, L21

\bibitem[Decressin et al.(2007)]{decressin01} Decressin, T., Meynet, G., Charbonnel, C., Prantzos, N., \& Ekstr{\"o}m, S.\ 2007, \aap, 464, 1029 

\bibitem[D'Ercole et al.(2016)]{dercole01} D'Ercole, A., D'Antona, F., Vesperini, E. 2016, MNRAS, 461, 4088

\bibitem[D'Ercole et al.(2008)]{dercole02} D’Ercole A., Vesperini E., D’Antona F., McMillan S. L. W., Recchi S. 2008, MNRAS, 391, 825

\bibitem[Eisenstein et al.(2011)]{eis11} Eisenstein, D.~J., Weinberg, D.~H., Agol, E. et al. 2011, \aj, 142, 72

\bibitem[Garc{\'{\i}}a-Hern{\'a}ndez et al.(2015)]{garcia03} Garc{\'{\i}}a-Hern{\'a}ndez, D.~A., M{\'e}sz{\'a}ros, S., 
	Monelli, M. et al. 2015, \apjl, 815, L4

\bibitem[Gratton et al.(2012)]{gra01} Gratton, R.~G., Carretta, E., \& Bragaglia, A. 2012, \aapr, 20, 50

\bibitem[Gratton et al.(1999)]{gra05} Gratton, R. G., Carretta, E., Eriksson, K., \& Gustafsson, B. 1999, A\&A, 350, 955

%\bibitem[Gratton et al.(2000)]{gra06} Gratton, R. G., Sneden, C., Carretta, E. \& Bragaglia, A. 2000, A\&A, 354, 169

\bibitem[Gunn et al.(2006)]{gunn01} Gunn, J.~E., Siegmund, W.~A., Mannery, E.~J. et al. 2006, AJ, 131, 2332


\bibitem[Ivans et al.(2001)]{ivans01} Ivans, I.~I., Kraft, R.~P., Sneden, C.~S., et al. 2001, \aj, 122, 1438

\bibitem[Ivans et al.(1999)]{ivans02} Ivans, I.~I., Sneden, C., Kraft, R.~P. et al. 1999, \aj, 118, 1273


\bibitem[Johnson et al.(2015)]{johnson03} Johnson, C. I., McDonald, I., Pilachowski, C. A., et al. 2015, AJ, 149, 71

\bibitem[Johnson \& Pilachowski(2012)]{johnson01} Johnson, C.~I., \& Pilachowski, C.~A. 2012, \apjl, 754, L38

\bibitem[Kraft(1994)]{kraft01} Kraft, R.~P. 1994, \pasp, 106, 553

\bibitem[Kraft et al.(1992)]{kraft02} Kraft, R.~P., Sneden, C., Langer, G.~E., \& Prosser, C.~F. 1992, \aj, 104, 645

\bibitem[Lai et al.(2011)]{lai02} Lai, D.~K., Smith, G.~H., Bolte, M. et al. 2011,  \aj, 141, 62

\bibitem[Lapenna et al.(2014)]{lapenna02} Lapenna, E. et al. 2014, \apj, 797, 124

\bibitem[Lapenna et al.(2016)]{lapenna01} Lapenna, E., Lardo, C., Mucciarelli, A. et al. 2016, \apjl, 826, L1

\bibitem[Lardo et al.(2012)]{lardo01} Lardo, C., Pancino, E., Mucciarelli, A. \& Milone, A.~P. 2012, \aap, 548, A107

\bibitem[Lardo et al.(2013)]{lardo03} Lardo, C. et al. 2013, \mnras, 433, 1941

\bibitem[Lardo et al.(2017)]{lardo02} Lardo, C., Salaris, M., Savino, A., Donati, P., Stetson, P.~B. 
		\& Cassisi, S. 2017, \mnras, 466, 3507

\bibitem[Majewski et al.(2017)]{majewski01} Majewski, S.R., Schiavon, R.~P., Frinchaboy, P.~M. et al. 2017, AJ, 154, 94 


%\bibitem[Marino et al.(2011)]{marino03} Marino, A. F. et al. 2011, A\&A, 532, 8

\bibitem[Marino et al.(2012)]{marino06} Marino, A. F., Milone, A. P., Piotto, G. et al. 2012, \apj, 746, 14

\bibitem[Marino et al.(2016)]{marino05} Marino, A. F., Milone, A. P., Casagrande, L. et al. 2016, \mnras, 459, 610

\bibitem[Marino et al.(2017)]{marino04} Marino, A.~F., Milone, A.~P., Yong, D. et al. 2017, \apj, 843, 66

\bibitem[M{\'e}sz{\'a}ros et al.(2015)]{meszaros01} M{\'e}sz{\'a}ros, S., Martell, S.~L., Shetrone, M. et al. 2015, \aj, 149, 153

\bibitem[Milone et al.(2008)]{milone01} Milone, A.~P., Bedin, L.~R., Piotto, G. et al.\ 2008, \apj, 673, 241 

\bibitem[Milone et al.(2013)]{milone04} Milone, A. P., et al. 2013, A\&A, 555, 143

\bibitem[Milone et al.(2015a)]{milone02} Milone, A.~P., Marino, A.~F., Piotto, G. et al. 2015, \mnras, 447, 927

\bibitem[Milone et al.(2015b)]{milone03} Milone A.~P. et al. 2015b, \apj, 808, 51

\bibitem[Milone et al.(2012)]{milone05} Milone, A.~P., Piotto, G., Bedin, L.~R. et al. 2012, \apj, 744, 58

\bibitem[Milone et al.(2017)]{milone06} Milone, A. P., Piotto, G., Renzini, A. et al. 2017, \mnras, 464, 3636

\bibitem[de Mink et al.(2009)]{demink01} de Mink, S.~E., Pols, O.~R., Langer, N., \& Izzard, R.~G.\ 2009, \aap, 507, L1 

\bibitem[Monelli et al.(2013)]{monelli01} Monelli, M., Milone, A.~P., Stetson, P.~B. et al. 2013, \mnras, 431, 2126

\bibitem[Nardiello et al.(2015)]{nardiello01} Nardiello D., Milone A. P., Piotto G., Marino A. F., Bellini A., Cassisi S.,
2015, A\&A, 573, A70


\bibitem[Nordlander \& Lind(2017)]{nord01} Nordlander, T. \& Lind, K. 2017, arXiv, 1708.01949

\bibitem[Pancino et al.(2017)]{pancino01} Pancino, E. et al. 2017, \aap, 601, 112

\bibitem[Piotto et al.(2007)]{piotto01} Piotto, G., Bedin, L.~R., Anderson, J., et al.\ 2007, \apjl, 661, L53 

\bibitem[Piotto et al.(2015)]{piotto02} Piotto,~G., Milone,~A.~P. , Bedin,~L.~R. et al. 2015, \aj, 149, 91

\bibitem[Ram{\'{\i}}rez \& Cohen(2003)]{ram01} Ram{\'{\i}}rez, S.~V. \& Cohen, J.~G. 2003, \aj, 125, 224

\bibitem[Renzini et al.(2015)]{renzini01} Renzini, A., D’Antona, F., Cassisi, S., et al. 2015, \mnras, 454, 4197

\bibitem[Roederer \& Sneden(2011)]{roederer01} Roederer, I.~U. \& Sneden, C. 2011, \aj, 142, 22

\bibitem[Sbordone et al.(2011)]{sbordone01} Sbordone, L., Salaris, M., Weiss, A. \& Cassisi, S. 2011, \aap, 534, 9

\bibitem[Smith et al.(1996)]{smith03} Smith, G.~H., Shetrone, M.~D., Bell, R.~A., Churchill, C.~W., 
		\& Briley, M.~M. 1996, \aj, 112, 1511

\bibitem[Sneden et al.(2000)]{sneden02} Sneden, C., Pilachowski, C.~A., \& Kraft, R.~P. 2000, \aj, 120, 1351

\bibitem[Sneden et al.(2004)]{sneden01} Sneden, C., Kraft, R.~P., Guhathakurta, P., Peterson, R.~C., \& Fulbright, J.~P. 2004, \aj, 127, 2162

\bibitem[Sneden et al.(1992)]{sneden03} Sneden, C., Kraft, R.~P., Prosser, C.~F., \& Langer, G.~E. 1992, \aj, 104, 2121

\bibitem[Sneden et al.(1997)]{sneden05} Sneden, C., Kraft, R.~P., Shetrone, M.~D. et al. 1997, \aj, 114, 1964

\bibitem[Sobeck et al.(2011)]{sobeck01} Sobeck, J.~S., Kraft, R.~P., Sneden, C. et al. 2011, \aj, 141, 175


\bibitem[Soto et al.(2017)]{soto01} Soto, M., Bellini, A., Anderson, J. et al. 2017, \aj, 153, 19

\bibitem[Stetson et al.(2014)]{stetson01} Stetson, P. B., Braga, V. F., Dall’Ora, M., et al. 2014, PASP, 126, 521

\bibitem[Wilson et al.(2012)]{wil10} Wilson, J., Hearty, F., Skrutskie, M.~F. et al. 2012, SPIE, 8446, 84460H

\bibitem[Ventura et al.(2016)]{ventura03} Ventura, P. et al. 2017, \apjl, 831, L17

\bibitem[Ventura et al.(2001)]{ventura01} Ventura, P., D'Antona, F., Mazzitelli, I., \& Gratton, R.\ 2001, \apjl, 550, L65 

\bibitem[Yong et al.(2009)]{yong04} Yong, D., Grundahl, F., D'Antona, F. et al. 2009, \apjl, 685, L62



\bibitem[Zhang et al.(2017)]{zhang01} Zhang, J., Shi, J., Pan, K., Allende Prieto, C., Liu, C. 2016, \apj, 833, 137
\end{document}